\documentstyle[prd,aps,floats,epsfig,graphicx]{revtex}
\input epsf
\def\pht{\tilde{\phi}}
\def\xt{\tilde{x}}
\def\pt{\partial}
\newcommand{\be}{\begin{equation}}
\newcommand{\ee}{\end{equation}}
\newcommand{\bea}{\begin{eqnarray}}
\newcommand{\eea}{\end{eqnarray}}
\begin{document}

\title{\bf{The Stress-Energy Tensor in Soluble Models of
Spherically Symmetric Charged Black Hole Evaporation}}
\author{Kamran Diba\thanks{
e-mail: diba@het.brown.edu} and David A. Lowe}
\address{Department of Physics, \\Brown University,
\\Providence,
RI 02912, \\U.S.A.}

\maketitle

\begin{abstract}
We study the decay of a near-extremal black hole in AdS$_2$,
related to the near-horizon region of
$3+1$-dimensional Reissner-Nordstr\"om spacetime, following
Fabbri, Navarro, and Navarro-Salas.
Back-reaction is included in a semiclassical approximation. Calculations of the stress-energy tensor of
matter coupled to the physical spacetime for an affine null
observer demonstrate that the black hole
evaporation proceeds smoothly and the near-extremal black hole
evolves back to an extremal ground state, until this approximation
breaks down.
\end{abstract}

\section{Introduction}

The fate of near-extremal black holes during quantum evaporation
has been of much interest because they present an excellent
laboratory for investigating the information paradox. These black
holes possess a stable ground state, namely the extremal black
hole, and are able to avoid some of the problems which plague
uncharged black holes during evaporation. For example, in the
well-studied linear dilaton black hole model of Callan, Giddings,
Harvey and Strominger (the so-called CGHS model) \cite{cghs}, as
the black hole evaporates away, the outer horizon encounters a
naked singularity \cite{lowefirst}.  Charged black holes, on the
other hand, possess a double horizon structure, with an inner and
outer apparent horizon.  In the extremal limit, where the black
hole mass approaches the black hole charge in the appropriate
units, the distance between the horizons is zero and they rest at
some finite value of the radius (the extremal radius).  The
singularity at the center of the black hole thus lies safely
behind the horizons, and there is no risk of encountering a naked
singularity, even at the endpoint of evaporation.  This fact, as
well as their frequent appearance in string theory, makes them
particularly appealing for investigation.

Jacobson has suggested that the semi-classical evolution of
near-extremal black holes may break down while still far from
extremality \cite{jacobson}. Using adiabatic arguments, Jacobson
claims that in-falling photons created at the outer apparent
horizon during Hawking evaporation will unavoidably fall through
the inner horizon as well.  If the photons encounter a large
buildup of energy behind the inner horizon then the inner horizon
is unstable and the semiclassical approximation is invalid.
Otherwise, the photons will eventually pile up behind the outer
horizon, causing it to become unstable. In either scenario, the
semiclassical approximation may break down long before one would
expect based on thermodynamic/statistical mechanics arguments
\cite{wilczetc}.

String theory, on the other hand, suggests that the evaporation
should proceed in a smooth way, with the excited black hole
returning to its ground state. In particular, extremal black holes
with Ramond-Ramond charge in Type II string theories are described
by configurations of D-branes, which have exact conformal field
theory descriptions. At least at weak string coupling, it can be
verified that the decay back to the extremal state is regular.

To investigate this issue, we study
the behavior of the stress-energy tensor for a
freely-falling observer during the evaporation process. We calculate this
and other observables for an extremal black hole after a shock
perturbation of mass is sent in.  In particular, we look for signs
of instability and energy buildup behind the inner and outer
horizons.  We will use the semi-classical approximation and verify
its validity close to the endpoint of evaporation, where it breaks
down according to the criterion of \cite{wilczetc}.
In this paper, we utilize a model set forth
by Fabbri, Navarro, and Navarro-Salas  \cite{nav1,nav2,navlong}.

The Reissner-Nordstr\"om black hole line element is
\bea
\label{4drn}
ds^2 &=& -\left(1-\frac{2l^2m}{r}+\frac{l^2q^2}{r^2}\right)dt^2 +
   \left(1-
   \frac{2l^2m}{r}+\frac{l^2q^2}{r^2}\right)^{-1}dr^2 +r^2 d\Omega^2\\
   &=& -\frac{(r-r_+)(r-r_-)}{r^2}dv^2 + 2 dr dv + r^2 d\Omega^2~,
\eea
with $r_{\pm}=l^2m \pm l\sqrt{l^2m^2-q^2}$.  Setting
$\phi=r^2/4l^2$ and $l=\sqrt{G_{N}}$, we can conformally rescale
the metric by $d\tilde{s}^2=\sqrt\phi ds^2$ and describe its
two-dimensional reduction using the following action:
\be\label{firstact}
S=\int d^2x \sqrt{-g} \left[R\phi + l^{-2}V(\phi) \right]\,,
\ee
where $V(\phi)=(4\phi)^{-1/2}-q^2(4\phi)^{-3/2}$. The extremal
black hole radius corresponds to when $V(\phi_0)=0$ \cite{bol},
giving $\phi_0=q^2/4$.  A trapped surface, in a metric of the form
\be
ds^2=g_{\mu\nu}dx^{\mu}dx^{\nu}+\phi^2d\Omega^2~,
\ee
corresponds to when the two-sphere at $x^+ ,x^-$ is decreasing in
both null directions: $\pt_{\pm}\phi<0$.  Asymptotically, in these
solutions, $\pt_-\phi<0$ and $\pt_+\phi>0$.  Therefore an
apparent horizon, which is the outer boundary of a trapped
region, occurs at $\pt_+\phi=0$.

The action above, (\ref{firstact}), can also be derived from the
Reissner-Nordstr\"om action considered in \cite{triv} and
\cite{trivstrom}
\be
S=\int d^2x\sqrt{-g}\left[e^{-2\phi}R+2e^{-2\phi}(\nabla\phi)^2 +
2 -
  2q^2e^{2\phi}\right]~,
\ee
which, in the conformal gauge ($ds^2=-e^{2\rho}dx^+dx^-$), is
equal to
\be
S=\int d^2x \left[ 4\pt_+\pt_-\rho e^{-2\phi}-4\pt_+\phi\pt_-\phi
e^{-2\phi} +
  e^{2\rho}-q^2e^{2\rho+2\phi} \right] \; .
\ee
By letting $\rho\rightarrow \rho+\phi/2$, which amounts to a
conformally rescaling of the metric by
$d\tilde{s}^2=e^{\phi}ds^2$, we can rewrite this as
\be
S=\int d^2x \left[4\pt_+\pt_-\rho e^{-2\phi}+e^{2\rho}e^{\phi}-
  q^2e^{2\rho}e^{3\phi} \right]\; ,
\ee
or
\be
S=\int d^2x \sqrt{-g} \left[Re^{-2\phi} + V(\phi) \right] \; ,
\ee
with $V(\phi)= 2e^{\phi} - 2q^2e^{3\phi}$.  Now redefining
$e^{-2\phi}$ as $\phi$ the action can be expressed as
\be
S=\int d^2x \sqrt{-g} \left[R\phi + V(\phi)\right] \; ,
\ee
with $V(\phi)=2/\phi^{1/2}-2q^2/\phi^{3/2}$, and $\phi=r^2$.  This
is equivalent to (\ref{firstaction}) with $l^2=1/4$ and $q^2$
rescaled to $2q^2$.

Returning to (\ref{firstact}), performing an expansion of $\phi$
around $\phi_0$ to first order ($\phi=\phi_0 + \pht$) in the
action yields an effective near-extremal action
\be
\label{firstaction}
S = \int d^2x \sqrt{-g} \left[R \tilde{\phi} + 4 \lambda^2
\tilde{\phi} \right]~.
\ee
We must keep that approximation in mind when making statements
derived from this action.

When a shock mass is added to the black hole mass at $v=v_0$,
$r_{\pm}$ in the metric (\ref{4drn}) becomes modified
\bea
r_{\pm}&=&l^2m + l^2\Delta m \pm l\sqrt{l^2(m+\Delta m)^2-q^2}\\
&\approx& l^2m \pm l\sqrt{l^2m^2 +2l^2m\Delta m -q^2}~,
\eea
to lowest order in $\Delta m$.  In the extremal case when $m=q/l$, we
get
\be
r_{\pm}=\pm lq \pm l^2\sqrt{2m\Delta m} \; .
\ee
Letting $r_0 = lq$, this translates into
\bea
ds^2 &=& -\frac{(r-r_0 -l\sqrt{2r_0\Delta m})(r-r_0 +l\sqrt{2r_0\Delta
    m})}{r_0^2}dv^2 +2ldrdv \\
&=& -\frac{(r-r_0)^2-2lr_0\Delta m}{r_0^2}dv^2 +2ldrdv \\
&=& -\left(\frac{\delta r^2}{r_0^2}-\frac{2l\Delta m}{r_0}\right)dv^2
+2ldrdv \\
&=& -\left(\frac{\pht^2}{q^4}-\frac{2\Delta m}{q}\right)dv^2
+2l\frac{d\phi}{\sqrt{\phi}}dv \; ,
\eea
which leads to
\be
\label{themetric}
d\tilde{s}^2=\sqrt{\phi}ds^2=-\left(2\frac{\pht^2}{q^3}-l
m_S(v)\right)dv^2 +2ld\pht dv \; ,
\ee
where $m_S(v)$ is the shock mass perturbing the black hole written
as a function of the null coordinate $v$ (and equal to zero for
$v<v_0$).

So far, we have been describing an ``eternal'' black hole. In
order to study the Hawking radiation of these black holes, we must
add dynamical matter fields to the action.  Here, this is done by
adding N minimally coupled scalar fields and studying the large N
limit where the one-loop quantum correction adequately describes
the effect of the Hawking radiation.  This may not correspond to
the most physically accurate way of describing the matter fields,
but it is the most calculationally simple \cite{balfab,gidds}.  For this coupling of the matter fields, the effect
of the back-reaction on the spacetime geometry can be
semiclassically included by adding a Liouville-Polyakov term
\cite{polyakov}.
\be\label{fullaction}
I = \int d^2x \sqrt{-g} \left[ R \tilde{\phi} + 4
\lambda^2 \tilde{\phi} -\frac{1}{2} \sum_{i=1}^N |\nabla f_i|^2 \right]
- \frac{N\hbar}{96\pi} \int d^2x \sqrt{-g} R  \Box^{-1} R + \xi
\frac{N\hbar}{12\pi} \int d^2x \sqrt{-g} \lambda^2 ~.
\ee
Working in the conformal gauge where
\be
\label{confmetric}
d\tilde{s}^2=-e^{2\rho}dx^+dx^- \; ,
\ee
the equations of motions become
\bea
2\pt_+\pt_-\rho+\lambda^2e^{2\rho}&=&0 \,, \\
\pt_+\pt_-\pht+\lambda^2e^{2\rho}\left(\pht +
(\xi-1)\frac{N\hbar}{12\pi}\right)&=&0 \, ,\\
\pt_+\pt_- f_i&=&0 \, , \\
\label{stresseqns}
-2\pt^2_{\pm} \pht +4\pt_{\pm}\rho\pt_{\pm} \pht
 &=& T^f_{\pm \pm}-\frac{N\hbar}{12\pi} \left((\pt_{\pm} \rho )^2
-\pt_{\pm}^2\rho +t_{\pm}(x^{\pm})\right) \; .
\eea
$\phi$ can always be shifted to absorb the $\xi-1$ term, so
without loss of generality, $\xi$ is set equal to $1$.  The entire
right hand side of the final equation represents the full
(classical plus quantum) matter stress-energy tensor. The
functions, $t_{\pm}(x^{\pm})$, are determined by the
boundary  conditions and depend on the vacuum choice.  Under
coordinate transformations, they transform according to
\be
\label{schtrans}
\left(\frac{dz'}{dz}\right)^2 t_{z'}(z') = t_z(z) -
\frac{1}{2}\{z',z\} \; ,
\ee
where $\{z',z\}$ is the Schwarzian derivative defined by
\be
\{f,z\}=\frac{2 \pt^3_z f \pt_z f - 3 \pt^2z f
  \pt^2_z f} {2 \pt_z f \pt_z f} \; .
\ee
The non-tensor transformation of the functions,
$t_{\pm}(x^{\pm})$, arises as a direct consequence of the
non-local nature of the Liouville-Polyakov term \cite{cghs,balfab}
.  Among the other terms which appear in (\ref{stresseqns}),
$T^f_{\pm \pm}$, which is the classical part of the total
stress-energy, transforms as a tensor.  $(\pt_{\pm} \rho )^2
-\pt_{\pm}^2\rho$, on the other hand, transforms according to the
Schwarzian transformation equation (\ref{schtrans}).  We can see
this by letting $x^+\rightarrow \xt^+$,
\be
e^{2\rho}dx^+dx^- =  e^{2\rho}\frac{dx^+}{d\xt^+}d\xt^+dx^- \; ,
\ee
i.e.
\be
\rho \rightarrow \tilde{\rho}=\rho+
\frac{1}{2}log\left(\frac{dx^+}{d\xt^+}\right) \; .
\ee
Plugging $\tilde{\rho}$ in, we can confirm that
\be
\left(\frac{d\tilde{z}}{dz}\right)^2 \left(
(\pt_{\tilde{z}}\tilde{\rho})^2-
  \pt_{\tilde{z}}^2\tilde{\rho}\right)= \left((\pt_{z} \rho
  )^2-\pt_{z}^2\rho\right) + \frac{1}{2}\{z',z\} \; ,
\ee
so that with the addition of the $t_{\pm}(x^{\pm})$ the entire
right-hand side of the stress-energy equations transforms as a
tensor under a change of coordinate.  As can be seen, this is also
consistent with how the left hand side transforms. The complete
matter stress-energy tensor, which we denote simply by $T_{\pm
\pm}$,
\be\label{stressdef}
T_{\pm \pm}=T^f_{\pm \pm}-\frac{N\hbar}{12\pi} \left( (\pt_{\pm} \rho )^2 -
  \pt_{\pm}^2\rho+t_{\pm}(x^{\pm})\right)=
-2\pt^2_{\pm} \pht +4\pt_{\pm}\rho\pt_{\pm} \pht~,
\ee
transforms simply as a tensor under coordinate transformations for
a given vacuum choice.

It is useful to define the vacuum in flat spacetime
\be\label{flat}
ds^2=\eta_{\mu\nu}dx^{\mu}dx^{\nu} \;.
\ee
The scalar fields, $f_i$, can be decomposed into
\be\label{modedecom}
f_i(x)=\sum_{j}\left[a_ju_j(x)+ a_j^{\dag}u_j^*(x)\right]~,
\ee
where
\be
u_j(x)=\frac{1}{\sqrt{4\pi\omega}}e^{ik\cdot x}~, \;\;\;\;
    k^0=\omega~,
\ee
form a complete orthonormal set.  The vacuum state, $|0\rangle$,
is then defined such that
\be\label{modedef}
a_j|0\rangle=0 \;, \;\;\;\; \forall \;j \;.
\ee
If we wish to work with a conformally related spacetime
\be
g_{\mu\nu}(x)=\Omega^2(x)\eta_{\mu\nu} \;,
\ee
the scalar fields transform according to
\be
f_i(x)=\Omega^{(2-n)/2}\sum_{j}\left[a_j u_j(x) +
    a_j^{\dag}u_j^*(x)\right] \;.
\ee
Now the vacuum state associated with the modes defined by
(\ref{modedef}) is known as the conformal vacuum.

In two dimensions, it is possible to express the stress-energy of
a spacetime conformally related to flat spacetime, $ds^2=dudv$, by
$d\bar{s}^2=C(u,v)dudv$ \cite{birdav}:
\be
\label{conftrans}
\langle T_\mu^\nu(g)\rangle=(-g)^{-1/2}\langle
T_\mu^\nu(\eta)\rangle
    + \theta_\mu^\nu -(1/48\pi)R\delta_\mu^\nu
\ee
where
\bea
\theta_{uu}&=&-(\hbar/12\pi)C^{1/2}\pt^2_u C^{-1/2}\\
\theta_{vv}&=&-(\hbar/12\pi)C^{1/2}\pt^2_v C^{-1/2}\\
\theta_{uv}&=&\theta_{vu}~,
\eea
for each scalar field.  These $\theta$ terms give the Schwarzian
derivatives of a function $h(u)$ when one makes the substitution
$C(u,v)=\partial h/\partial u$.  If the state used in evaluating
the expectation value in flat spacetime is a vacuum state, then
the state appearing in the curved spacetime expectation value is
referred to as a conformal vacuum.

It is also possible to relate the stress-energy tensors defined in
different flat spacetimes, $d\bar{s}^2=d\bar{u}dv$ and $ds^2=dudv$
\cite{davies,birdav}. Following \cite{davies}, consider a general
metric
\be
ds^2=C(u,v)dudv \; ,
\ee
with
\be
\label{cmetric}
C(u,v)=A(\bar{u},\bar{v})\frac{d\bar{u}}{du}\frac{d\bar{v}}{dv}~,
\ee
where
\bea
v&=&\beta(\bar{v})\\
u&=&\beta(\bar{u}-2R_0) \; .
\eea
Using (\ref{conftrans}), the stress-energy tensor with respect to
the conformal vacuum is given by
\bea
T_{uu}&=&-F_u(C)\\
T_{vv}&=&-F_v(C) \; ,
\eea
where $F$ denotes the function
\be
F_x(y)\equiv(12\pi)^{-1}y^{1/2}\pt^2_x y^{-1/2} \; .
\ee
For (\ref{cmetric}) in $\bar{u},\bar{v}$ coordinates
\bea
T_{\bar{u}\bar{u}}&=&-F_{\bar{u}}(A)+F_{\bar{u}}(\beta ') \;\;\; \beta=\beta(\bar{u}-2R_0) \\
T_{\bar{v}\bar{v}}&=&-F_{\bar{v}}(A)+F_{\bar{v}}(\beta ') \;\;\;
\beta=\beta(\bar{v})\; .
\eea
However, since the first term on the right hand side is equivalent
to the stress-energy with respect to the conformal vacuum of
$A(\bar{u},\bar{v})d\bar{u}d\bar{v}$, this relation allows us to
relate the stress-energy tensors expressed with respect to two
different vacua.  It can be summarized as
\be
\label{vactrans}
\left(\frac{du}{d\bar{u}}\right)^2\langle 0|T_{uu}|0\rangle =
        \langle \bar{0}|T_{\bar{u}\bar{u}}|\bar{0}\rangle
        -\frac{N\hbar}{24\pi}\{u,\bar{u}\}\; .
\ee
Note that the last term on the right of equation (\ref{vactrans})
corresponds to the transformation of the stress-energy tensor when
the conformal factor is $d\bar{u}/du$.  That is to say, if we
define the vacuum state with respect to the positive energy modes
decomposed in $d\bar{s}^2=d\bar{u}dv$ and transform to a
conformally related spacetime, $ds^2 = dudv =
\frac{du}{d\bar{u}}d\bar{u}dv$, we obtain equation
(\ref{vactrans}).

Let us see now what happens when we express (\ref{themetric}) in
null coordinates.  The coordinate transformation $u=v+lq^3/\pht$
puts the metric into the form
\be
\label{lightcone}
d\tilde{s}^2=-2\frac{\pht^2}{q^3}du dv~,
\ee
for $v<v_0$, and
\be\label{vmet}
d\tilde{s}^2=-\left(2\frac{\pht^2}{q^3}-l\Delta m\right)
    d\bar{u} dv~,
\ee
for $v>v_0$ with
\be
\bar{u}=v+\sqrt{\frac{2lq^3}{\Delta
    m}}{\rm arctanh} \left(\sqrt{\frac{2}{lq^3\Delta
  m}} \pht \right) \; .
\ee
Now we can see the relation between $\bar{u}$ and $u$:
\be
u=v + \sqrt{\frac{2lq^3}{\Delta
 m}}{\rm cotanh}\left(\sqrt{\frac{\Delta m}{2lq^3}}(\bar{u}-v) \right)
 \; .
\ee
Thus the outgoing flux in the null coordinate $\bar{u}$ can be
calculated, since it is known that $T_{uu}=0$ before the shock
mass is introduced.  Therefore
\be
\label{outflux}
T_{\bar{u}\bar{u}}=\frac{N\hbar}{24\pi}\{\bar{u},u\}=\frac{N\hbar}{24\pi
 lq^3}\Delta m~,
\ee
which is a constant Hawking flux of radiation.  We have not yet
considered the effects of the back-reaction, which will be done in
the subsequent section.  However, this preliminary examination
demonstrates that we have indeed an evaporating black hole.
\section{Solutions}
The general solution to the stated equations of motion can be
written in terms of four  chiral functions, $A_{\pm}(x^{\pm})$,
and $a_{\pm}(x^{\pm})$, \cite{chiralsol1,chiralsol2} with
\be
d\tilde{s}^2 = - \frac{\pt_+A_+
  \pt_-A_-}{(1+\frac{\lambda^2}{2}A_+A_-)^2} dx^+dx^-~,
\ee
and
\be
\pht=-\frac{1}{2}\left(\frac{\pt_+a_+}{\pt_+A_+}+\frac{\pt_-a_-}{\pt_-A_-}\right)+
\frac{\lambda^2}{2} \frac{A_+a_- +A_-a_+}{1+\frac{\lambda^2}{2}A_+A_-}~,
\ee
constrained by
\bea
\pt^2_{+}\left(\frac{\pt_{+}a_{+}}{\pt_{+}A_{+}}\right) -
\frac{\pt^2_{+}A_{+}}{\pt_{+}A_{+}}\pt_{+}\left(\frac{\pt_{+}a_{+}}
  {\pt_{+}A_{+}}\right)
&=&  T_{+ +} \\
\pt^2_{-}\left(\frac{\pt_{-}a_{-}}{\pt_{-}A_{-}}\right) -
\frac{\pt^2_{-}A_{-}}{\pt_{-}A_{-}}\pt_{-}\left(\frac{\pt_{-}a_{-}}
  {\pt_{-}A_{-}}\right)
&=&  T_{- -} \; .
\eea
The first case initially studied by Fabbri, Navarro, and
Navarro-Salas \cite{nav1}  consists of a shock mass, $\Delta m$,
sent into the extremal  black hole,
\be
\label{shockmetric}
ds^2=-\left(\frac{2\pht^2}{q^3}-l\Delta m \Theta(v-v_0)\right)dv^2
+2ld\phi dv \; .
\ee
The gauge choice of $A_+ = x^+$ and $A_- =
\frac{-2}{\lambda^2x^-}$, with $\lambda^2=l^{-2}q^{-3}$  yields
\be\label{rho}
e^{2\rho}=\frac{2l^2q^3}{(x^- -x^+)^2} \; ,
\ee
\be\label{gaugemetric}
ds^2 = -\frac{2l^2q^3}{(x^- -x^+)^2}dx^+dx^- \; .
\ee
This gauge fixes $((\pt_{\pm} \rho )^2 -\pt_{\pm}^2\rho)$ in the
constraint equations to be identically zero everywhere.  Thus,
$t_{\pm}(x^{\pm})$ represents the only quantum part of the
stress-energy tensor.  That is,
\be
T_{\pm \pm}=T^f_{\pm \pm}-\frac{N\hbar}{12\pi}
  t_{\pm}(x^{\pm})~.
\ee
An important consequence of this result is that the quantum nature
of the solutions only manifests itself in the boundary conditions.
The same solutions are obtained classically if the flux sent into
the black hole coincides with the quantum boundary conditions.
This will be discussed in more detail later on.  The gauge choice
itself corresponds to AdS$_2$ spacetime, with the AdS boundary
occurring at the coordinate singularity $x^-=x^+$. The metric
(\ref{shockmetric}) can be brought into the gauge-fixed form by
setting
\bea
a_+&=&-lq^3 \\
a_-&=&0~,
\eea
for $v<v_0$.  Requiring continuity at $v=v_0$, for $v>v_0$
\bea
a_+&=&-\frac{1}{2}\Delta mx^+_0(x^+-x^+_0)-lq^3 \; ,\\
a_-&=&l^2q^3\Delta m\frac{x^+_0}{x^-}-l^2q^3\Delta m \; .
\eea
Thus we have for $v<v_0$
\be\label{phivac}
\pht=\frac{lq^3}{x^- - x^+} \; , \;\;\;\;\; v=x^+ \; ,
\ee
and for $v>v_0$
\be
\label{phifirst}
\pht=lq^3\frac{1-\frac{\Delta
    m}{2lq^3}(x^+-x^+_0)(x^--x^+_0)}{x^--x^+} \; ,
\ee\be
\label{vcoord}
v =x^+_0+\sqrt{\frac{2lq^3}{\Delta
    m}}{\rm arctanh}\left[\sqrt{\frac{2lq^3}{\Delta
    m}}(x^+-x^+_0)\right]\; .
\ee
These solutions break down as we approach $x^- -x^+ \sim 4lq$,
since the small $\pht$ approximations then becomes invalid. There
is also a coordinate singularity in the metric that occurs when
$x^-=x^+$.   (\ref{phivac}) represents the vacuum of the
solutions.  The extremal radius, $\pht=0$, occurs at $x^--x^+
\rightarrow \infty$. In the region below the AdS boundary, $x^- <
x^+$, $\pht<0$.  That is, this region corresponds in fact to the
area behind the extremal black hole radius.  The double-horizon
structure manifests itself when we solve $\pt_+\pht=0$, giving a
horizon at $x^--x^+ \rightarrow \pm \infty$.  So $x^- > x^+$ and
$x^- < x^+$ correspond to two different coordinate patches of the
solutions, with $x^- < x^+$ corresponding to an area that actually
lies behind $x^--x^+ = \infty$.  In analyzing these results, we
study the area above the AdS boundary, $x^- > x^+$.

\begin{figure}[htbp]
\includegraphics[width=3.5in,angle=-90]{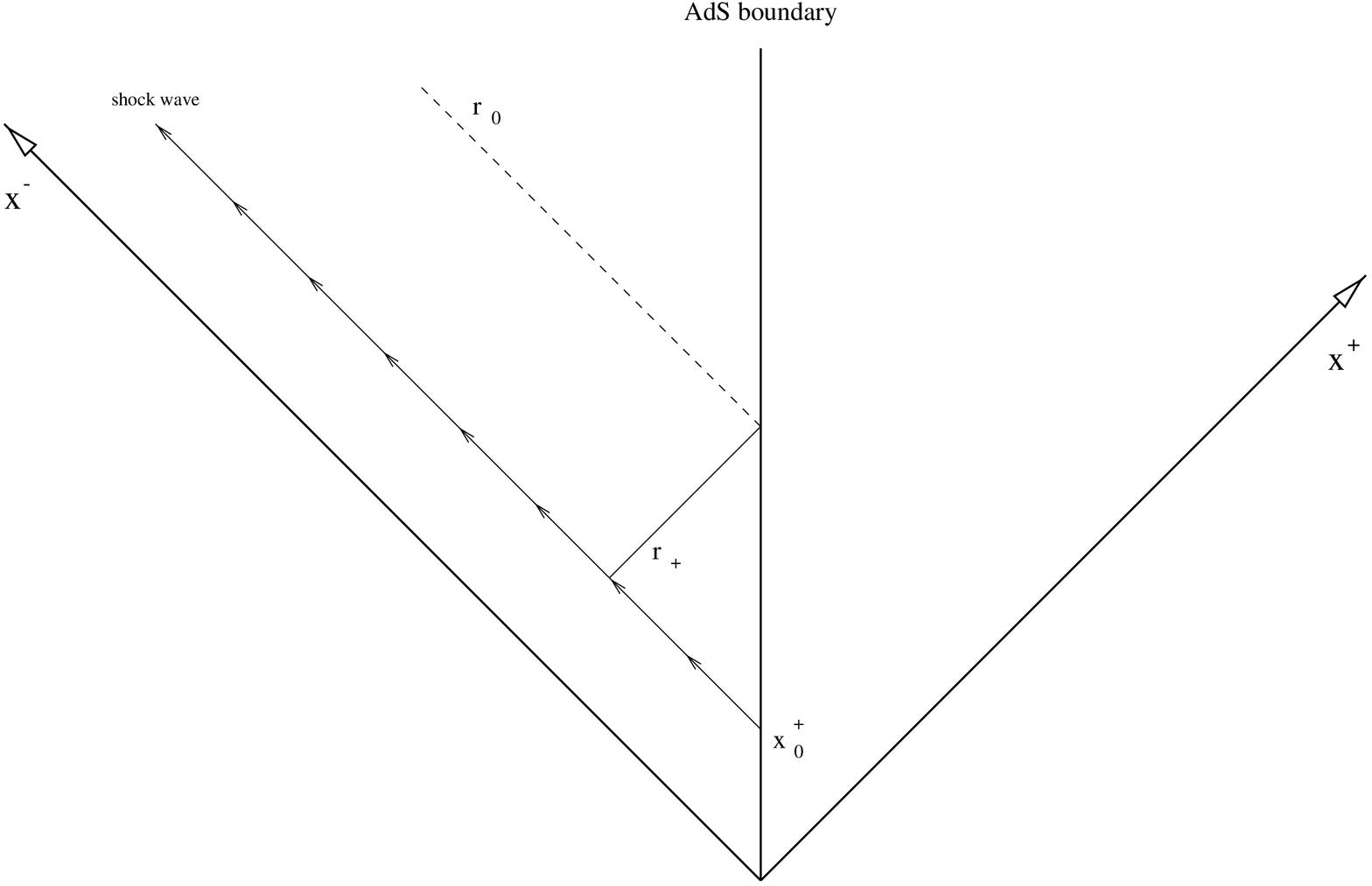}
\caption{Kruskal diagram for the static classical solution of the
near-extremal  RN  black hole. The AdS$_2$ boundary is seen at
$x^-=x^+$. $r_+$  represents the outer horizon, while $r_-$, the
inner horizon, lies at $x^+ \rightarrow \infty$.  $r_0$, the
extremal radius, does not meet with $r_+$ except at the AdS
boundary, which represents timelike infinity.}
\label{kruskal}
\end{figure}

(\ref{phifirst}) represents the classical solution shown in
Fig.~\ref{kruskal}. Let us first consider this case: the extremal
radius, $\pht=0$, occurs at $(x^+-x^+_0)(x^--x^+_0)=2lq^3/\Delta
m$. The apparent horizons, $\partial_+ \phi=0$, are given by
$x^-=x^+_0\pm \sqrt{2lq^3/\Delta m}$.  The AdS boundary represents
spatial infinity, i.e. the region infinitely far away from the
black hole where the radial variable $\phi$ becomes infinitely
large, with the black hole itself lying above $x^-> x^+$. We can
see then, that $r_+$ moves further out from the center of the
black hole for larger values of the  shock mass $\Delta m$, as
expected. $r_-$ can be understood to be at $x^- \rightarrow
\infty$.  $r_0$ and $r_+$ never meet (the apparent meeting point
is actually at infinity), as one would expect without Hawking
evaporation due to quantum effects.

We now wish to consider the semiclassical solutions.  The key to
solving these is picking the appropriate boundary conditions in a
given vacuum state.  The boundary condition is determined by the
behavior of the stress-energy flux at regions far outside of the
black hole, where statements can be made about the expected flux.
This amounts to making an appropriate choice for
$t_{\pm}(x^{\pm})$, as it represents the quantum part of the
stress-energy tensor, $T_{++}$ (recall that the conformal term
involving $\rho$ derivatives is zero in this gauge).

Fabbri, Navarro, and Navarro-Salas choose vacuum conformal to
spacetime
\be\label{vacspace}
ds^2=-e^{2\rho}dvdx^- = -e^{2\rho}\frac{dv}{dx^+}dx^+dx^-~,
\ee
with $v(x^+)$ given by equation (\ref{vcoord}).  That is, the mode
decompositions discussed beginning with (\ref{modedecom}) are with
respect to the flat spacetime, $dvdx^-$.  This vacuum choice is
not well explained in \cite{nav1,nav2,navlong}. It is stated there
that since in (\ref{lightcone}), before the shock mass is
introduced, the stress-energy is zero, then this represents the
natural vacuum choice. However, a problem arises because while
$v(x^+)$ is from (\ref{vcoord}), after the shock mass ($v>v_0$),
in order for
\be
ds^2=-\left(2\frac{\pht^2}{q^3}-l\Delta m\Theta(v-v_0)\right)dv du~,
\ee
to be put into the form (\ref{confmetric}), $u$ will also be a
non-trivial function of $x^-$.   In fact, it is given by
\be
u=v_0 + \sqrt{\frac{2lq^3}{\Delta m}}{\rm cotanh}
  \left(\sqrt{\frac{\Delta m}{2lq^3}}(x^--v_0) \right) \;.
\ee
Regardless, in arriving at (\ref{vacspace}), the vacuum choice of
\cite{nav1,nav2,navlong}, necessarily requires $u=x^-$. Therefore,
the choice of (\ref{vacspace}) as vacuum space is not consistent
with (\ref{lightcone}).

In general, choosing $v=v(x^+)$ is a rather restrictive condition
on these solutions. This can be seen by closer inspection: let us
require that $v=v(x^+)$, i.e. $dv/dx^+=f(x^+)$, or more
conveniently,
\be\label{dvdxp}
\frac{dv}{dx^+}=\frac{lq^3}{F(x^+)} \; .
\ee
To bring the metric (\ref{shockmetric}) to the form
(\ref{gaugemetric}) it is also necessary to require that
\be\label{partialmphi}
\pt_-\pht=-\frac{F(x^+)}{(x^--x^+)^2} \; ,
\ee
which means
\be
\pht=\frac{F(x^+)}{x^--x^+}+G(x^+) \; .
\ee
Plugging this into the equation of motion for $\pht$ then leads to
$G(x^+)=F'(x^+)/2$, so that
\be
\pht=\frac{F(x^+)}{x^--x^+}+\frac{F'(x^+)}{2} \; .
\ee
Using this form of $\pht$ in the equation (\ref{stressdef}) we see
that $T_{--}$ will be zero everywhere in the solutions! Therefore,
we realize now that we are confined to boundary conditions that
give $t_-(x^-)=0$.

As long as $v=v(x^+)$, then, the vacuum choice used corresponds to
zero quantum flux in the spacetime of (\ref{vacspace}), which,
while not of imperative physical interest, allows us to solve the
equations of motion with relative ease.
\bea
t_-(x^-) &=& 0 \; , \\
\label{tplus}
t_+(x^+) &=& \frac{1}{2}\{v,x^+\}=\frac{2lq^3\Delta m}{(2lq^3
    - (x^+-x^+_0)^2 \Delta m)^2} \; .
\eea
Again, there is no outgoing flux, $T_{--}$, being emitted from the
black hole in the $x^+$ direction.  The evaporation of the black
hole proceeds simply through the negative flux entering
the black hole.  This means that in a sense the evaporation of the
black hole is built into the solutions from the boundary
conditions.  As mentioned before, the classical solutions would
have yielded the same result, given the same negative ingoing
flux.  The contribution of (\ref{tplus}) to the stress-energy
tensor is
\be
T_{++}=-\frac{N\hbar}{12\pi}\frac{2lq^3\Delta
    m}{(2lq^3 - (x^+-x^+_0)^2 \Delta m)^2} \; ,
\ee
which does in fact correspond to a negative flux of energy that
increases as $x^+-x^+_0 \rightarrow \sqrt{2lq^3/\Delta m}$.
\bea
\label{apeqn}
\pt^3_+a_+=-\frac{N\hbar}{12\pi}t_+(x^+)&=&-\frac{N\hbar}{12\pi}\frac{2lq^3\Delta
    m}{(2lq^3 - (x^+-x^+_0)^2 \Delta m)^2} \; ,\\
6\pt_-a_-+6x^-\pt^2_-a_-+(x^-)^2\pt^3_-a_-&=&0 \; ,
\eea
can be integrated to solve for $a_+$ and $a_-$ by requiring
continuity of $\phi$ at $x^+=x^+_0$, and by putting
(\ref{shockmetric}) in the form (\ref{confmetric}). The general
solutions to (\ref{apeqn}) are
\bea
a_+&=&-\frac{1}{2}\Delta m x^+_0(x^+-x^+_0)-lq^3 +\frac{N\hbar}{\pi}P(x^+)\; .\,\\
a_-&=&l^2q^3\Delta m\frac{x^+_0}{x^-}-l^2q^3\Delta m \; ,
\eea
with
\be\label{Pofx}
P(x^+)=\frac{(x^+-x^+_0)}{48}-\frac{\frac{2lq^3}{\Delta m}-(x^+
    -x^+_0)^2}{48\sqrt{\frac{2lq^3} {\Delta m}}}{\rm arctanh}\left((x^+ -x^+_0)
    \sqrt{\frac{\Delta m}{2lq^3}} \right) \; .
\ee
The resulting solution for $\pht$ is
\be
\label{phitilde}
\pht= lq^3\frac{1-\frac{\Delta
m(x^+-x^+_0)(x^--x^+_0)}{2lq^3}}{x^- -
  x^+} + \frac{N \hbar}{\pi}\frac{P(x^+)}{x^- - x^+} +\frac{N
  \hbar}{2\pi}P'(x^+) \; .
\ee
Let us take a moment to carefully examine $P(x^+)$ and its
properties.   We may take note that ${\rm arctanh}(x)$ becomes
logarithmically  divergent as $x \rightarrow 1$.  However,
$(x^2-1){\rm arctanh}(x) \rightarrow 0$  as $x \rightarrow 1$
remains finite.  Therefore $P(x^+)$ becomes indeterminate for $x^+
-x^+_0 \ge \sqrt{2lq^3/\Delta m}$ and $P'(x^+)$, which is
logarithmically divergent, blows up. Since $\phi$ represents the
radial coordinate, and it depends directly on $P'(x^+)$, we can
interpret $x^+ -x^+_0 \rightarrow  \sqrt{2lq^3/\Delta m}$ as
spatially being infinitely far from the black hole. It is not
possible to evolve the solutions beyond this point. However, when
the inner and outer horizon meet again at the extremal radius, at
the endpoint of black hole evaporation, the semiclassical
approximation has already broken down.  This happens before we
reach the divergence of $P'(x^+)$.

\begin{figure}[htbp]
\includegraphics[width=3.5in,angle=-90]{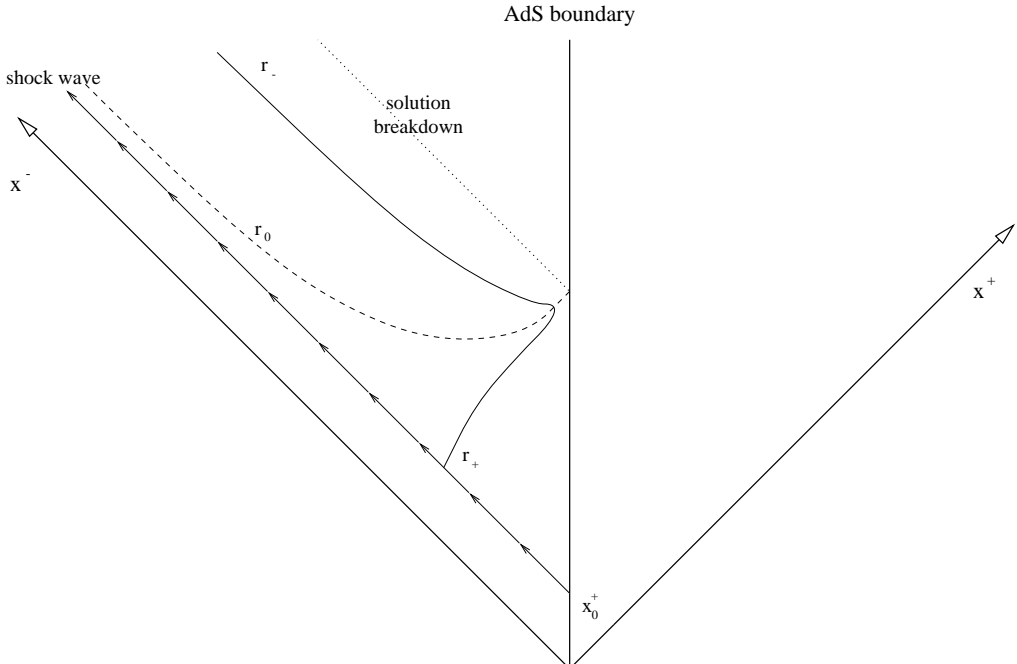}
\caption{Kruskal diagram for semiclassical solution of the
  near-extremal  RN black hole. The AdS$_2$ boundary is again seen at
  $x^-=x^+$. $r_+$ and $r_-$, which are given by $\pt_+\phi=0$
  evolve to meet at $r_0$ ($\phi_0$), the extremal radius.
  Close to this point, the semiclassical approximation breaks down.
  In addition, the solutions become indeterminate as $x^+-x^+_0 \rightarrow \sqrt{
  2lq^3/\Delta m}$.  The size of the shock mass $\Delta m$ determines the degree
  to which the outer horizon moves out from $x^- =\infty$.}
\label{finite}
\end{figure}

In the semiclassical solution, the inner and outer horizon come
together and meet at the extremal $r=r_0$ radius (see
Fig.~\ref{finite}), consistent with our picture of black hole
evaporation.  If we consider the extremal black hole as the limit
of a near-extremal black hole, it has a double-horizon which
becomes spatially separated with the introduction of a shock mass.
A larger shock mass corresponds to a bigger separation of $r_+$
and $r_-$.  However, as the black hole evaporates, the two
apparent horizons should eventually approach each other and return
to the extremal limit.  The classical solution of the equations of
motion (\ref{phifirst}) is recovered by taking the limit $\hbar
\rightarrow 0$.  These solutions do not demonstrate any outgoing
flux, as a result of the imposed condition that $T_{--}=0$.  All
evaporation manifests itself in a negative ingoing flux
$T_{++}<0$.  We should also keep in mind that (\ref{confmetric})
is conformally related to the dimensional reduction of the
physical metric (\ref{4drn}), so it is necessary to make further
calculations to understand what is really happening.

Note that there are approximations that have been made which need
to be re-examined.  Bringing the metric (\ref{shockmetric}) into
the conformal gauge form (\ref{confmetric}) necessarily requires
that
\be
\label{err1}
2l\pt_-\pht\pt_+v(x^+)+e^{2\rho}\ll 1
\ee
and
\be
\label{err2}
\left(-2\frac{\pht^2}{q^3}+l\Delta m\right)(\pt_+v(x^+))^2+2l\pt_+
   \pht\pt_+v(x^+)\ll 1 \; .
\ee
Putting solutions for $\pht$, (\ref{phitilde}), into (\ref{err1})
yields the constraint that
\be
\frac{4l^2q^3\kappa P(x^+)}{(x^--x^+)^2((x^+-x^+_0)^2\Delta
m-2lq^3)}\ll 1
\ee
plus a more complicated constraint that we omit due to space and
aesthetic considerations.  We must monitor the quantities on the
left hand side of the above equations to ensure that our
approximations are valid in the regions of interest.  This has
been done for all ensuing discussion.
\section{The Physical Metric}
The analysis has thus far been incomplete because the
stress-energy tensor considered does not correspond to an observer
living in $3+1$ dimensions. One of our new contributions is to study
components of the stress energy measured by an observer freely-falling
in the $3+1$-dimensional spacetime. This physical metric $ds^2$ is related to
the metric studied in the previous section $d\tilde s^2$ (\ref{confmetric}) by
the conformal factor, $\sqrt{\phi}$:
\be
\label{physmet}
ds^2=\frac{1}{\sqrt{\phi}}d\tilde{s}^2=-\frac{e^{2\rho}}{\sqrt{\phi}}dx^+
dx^- \; .
\ee
To construct a stress energy tensor that couples to this metric (\ref{physmet}) we
must add extra matter fields to the
action (\ref{fullaction}) that couple in a covariant way.
This addition will not alter the previous
equations of motion as long as the number of other matter fields $N$ is large.

It is relevant to consider what happens to a freely falling
observer coming in from far outside of the black hole.  Since it
is difficult to analytically describe the geodesic for an affinely
parameterized freely falling observer, we consider the next best
thing: a null in-falling observer. The Christoffel symbols for the
physical metric (\ref{physmet}) are
\bea
\Gamma^+_{++}=2\pt_+ \rho - \frac{1}{2}\pt_+ \log \phi \;,\\
\Gamma^-_{--}=2\pt_- \rho - \frac{1}{2}\pt_- \log \phi \;.\\
\eea
From the geodesic equation for $x^-$, we get
\be\label{geoeqn}
\frac{d^2 x^-}{d \tau^2} + (2\pt_- \rho - \frac{1}{2}\pt_- \log \phi)
        \frac{dx^-} {d \tau} \frac{dx^-}{d \tau} = 0 \; ,
\ee
with a similar equation for  $x^+$.  We consider the case where
$\tau$ is an  affinely parameterized null geodesic $\xt^-$ such
that
\be
ds^2= - \frac{e^{2\rho}}{\sqrt{\phi}} \frac{dx^+}{d \xt^-}\frac{d
x^-} {d \xt^-} = 0 \;.
\ee
This has solutions for fixed $x^+$, leaving $\xt^-=\xt^-(x^-)$.  Using
\be
\frac{\pt \rho}{\pt x^-}\frac{d x^-}{d \xt^-} = -
\frac{\pt \rho}{\pt x^+}\frac{d x^+}{d \xt^-} + \frac{d
  \rho} {d \xt^-}
\ee
and the fact that we are working with a null geodesic, the
equation (\ref{geoeqn}) reduces to
\be
\frac{d^2x^-}{d\xt^{-2}} + \frac{d \left(2\rho -
    \frac{1}{2}\log\phi\right) } {d \xt^-} \frac{dx^-}{d \xt^-}=0\;,
\ee
which is then solved to give
\be
\frac{d\xt^-}{d x^-}= C \frac{e^{2\rho(x^-)}}{\sqrt{\phi}} \;,
\ee
where $C$ is a constant of integration which we can set equal to
$1$. Using the conformal factor arising from (\ref{physmet}) in
the relation (\ref{conftrans}) gives
\be
T_{\pm\pm}=\frac{\hbar}{24}\left(\frac{3}{8}\frac{(\pt_{\pm}\phi)^2}{\phi^2}-
        \frac{1}{2}\frac{\pt^2_{\pm}\phi}{\phi}+\frac{\pt_+\rho\pt_+\phi}{\phi}+
        2\pt^2_+\rho-2(\pt_+\rho)^2-2t_{\pm}(x^{\pm})\right) \;,
\ee
where again, $t_{\pm}(x^{\pm})$ are determined by the boundary
conditions (i.e. vacuum choice).  Part of this tensor,
$N\hbar/12\pi(\pt^2_+\rho-(\pt_+\rho)^2-t_{\pm})$, is the source
term on the right hand side of the equations of motion. This is
because the matter fields couple to $\sqrt{-g}=e^{2\rho}/2$, and
hence the above terms contribute to the back-reaction.  The
stress-energy above is not a source for the back-reaction,
but is what an observer traveling through the physical spacetime
would measure.

In the affinely parameterized coordinates, the stress-energy is
\bea
\label{physstress}
\tilde{T}_{\pm\pm}&=&\left(\frac{dx^{\pm}}{d\xt^{\pm}}\right)^2T_{\pm\pm} \\
        &=&\frac{\hbar}{24}\frac{\phi}{e^{4\rho}}\left(\frac{3}{8}
        \frac{(\pt_{\pm}\phi)^2}{\phi^2}-\frac{1}{2}\frac{\pt^2_{\pm}\phi}
        {\phi}+\frac{\pt_+\rho\pt_+\phi}{\phi}+
        2\pt^2_+\rho-2(\pt_+\rho)^2-2t_{\pm}(x^{\pm})\right) \;,
\eea
where $\tilde{T}$ is used to denote the stress-energy tensor with
respect to $\xt^{\pm}$.
\begin{figure}[htbp]
\includegraphics[width=3.5in]{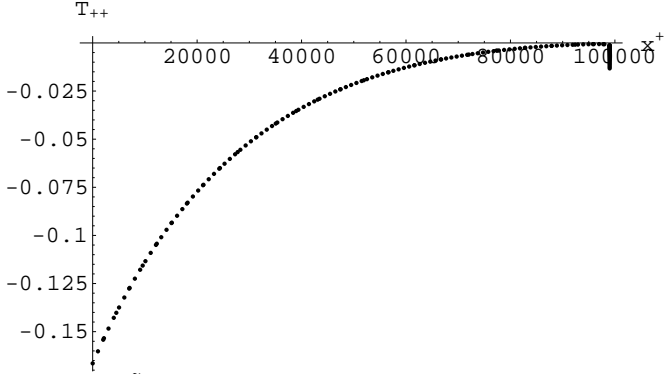}
\caption{$\tilde{T}_{++}$, the stress-energy of an affine null
  observer in physical spacetime
  $ds^2=-(e^{2\rho}/\sqrt{\phi})dx^+dx^-$,
  evaluated at points along the outer horizon, up to the point at which
  the two horizons meet again.  Notably, the evolution of the stress is smooth.
  In this vacuum choice, $\tilde{T}_{++}$ increases quickly with increasing $x^-$.
  Therefore, as the outer horizon recedes to meet the inner horizon, there is a large
  increase in the stress-energy toward the end of the evaporation.
  [$l=1,q=100;\Delta m=.0002; N\hbar=25\pi;\phi_0=2500$]}
\label{tppouter}
\end{figure}
\begin{figure}[htbp]
\includegraphics[width=3.5in]{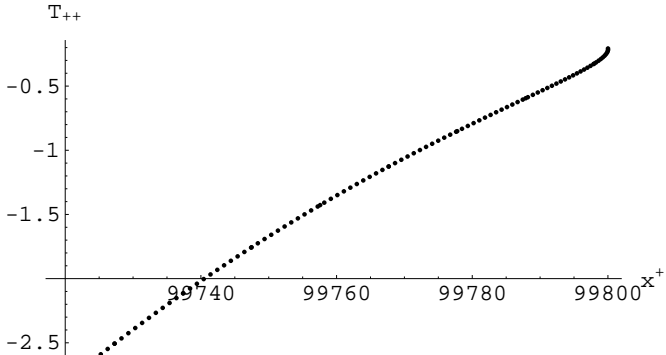}
\caption{$\tilde{T}_{++}$, the stress-energy of an affine null
  observer in physical spacetime $ds^2=-(e^{2\rho}/\sqrt{\phi})dx^+dx^-$,
  evaluated at points along the inner horizon, up to the point at which the
  two horizons meet again.  $\tilde{T}_{++}$ decreases very quickly with $x^-$, as
  the inner horizon moves on a nearly null-like trajectory.  [$l=1; q=100;
  \Delta m=.0002;N\hbar=25\pi;\phi_0=2500$]}
\label{tppinner}
\end{figure}
\begin{figure}[htbp]
\includegraphics[width=3.5in]{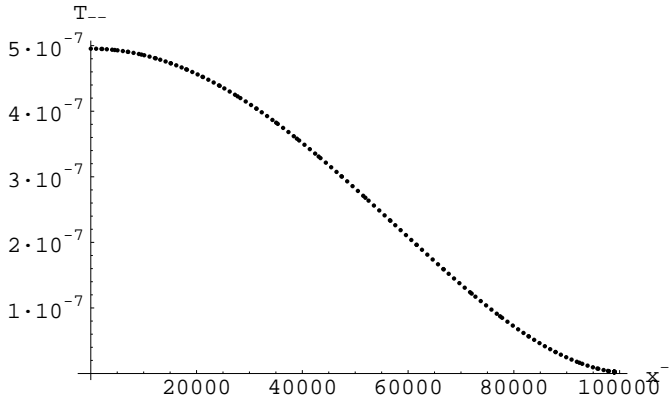}
\caption{$\tilde{T}_{--}$, the stress-energy of an affine null
  observer in physical spacetime $ds^2=-(e^{2\rho}/\sqrt{\phi})dx^+dx^-$,
  evaluated at points along the outer horizon, up to the point at which the
  two horizons meet again.  The behavior of $\tilde{T}_{--}$ reflects the Hawking
  radiation leaving the black hole and reaching zero as the black hole returns to
  extremality.  Along the inner horizon, $\tilde{T}_{--}$ is essentially zero.
  [$l=1,q=100;\Delta m=.0002; N\hbar=25\pi;\phi_0=2500$]}
\label{tmmouter}
\end{figure}
\begin{figure}[htbp]
\includegraphics[width=3.5in]{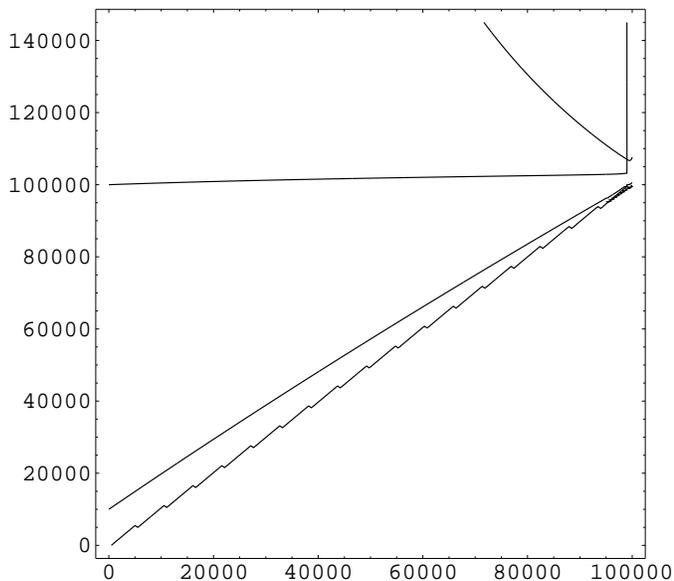}
\caption{The outer horizon is shown as it recedes to meet
  the inner horizon, at $\phi_0$.  Also shown is the contour for fixed radius
  $\pht=100$, near the AdS boundary, for which subsequent plots were made.
  [$l=1,q=100;\Delta m=.0002; N\hbar=25\pi;\phi_0=2500$]}
\label{phicontour}
\end{figure}
\begin{figure}[htbp]
\includegraphics[width=3.5in]{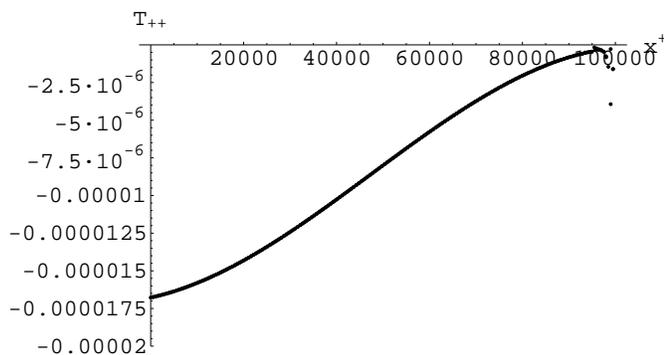}
\caption{$\tilde{T}_{++}$, the stress-energy of an affine null
  observer in physical spacetime $ds^2=-(e^{2\rho}/\sqrt{\phi})dx^+dx^-$,
  shown at points along the fixed radius
  $\pht =100$.  $\tilde{T}_{++}$ is negative, a remnant of the boundary
  conditions which demonstrates that negative flux gets sent in to reduce
  the black hole mass.  This flux goes to zero as the black hole returns to
  extremality.  [$l=1,q=100;\Delta m=.0002;
  N\hbar=25\pi;\phi_0=2500$]}
\label{tpp}
\end{figure}
\begin{figure}[htbp]
\includegraphics[width=3.5in]{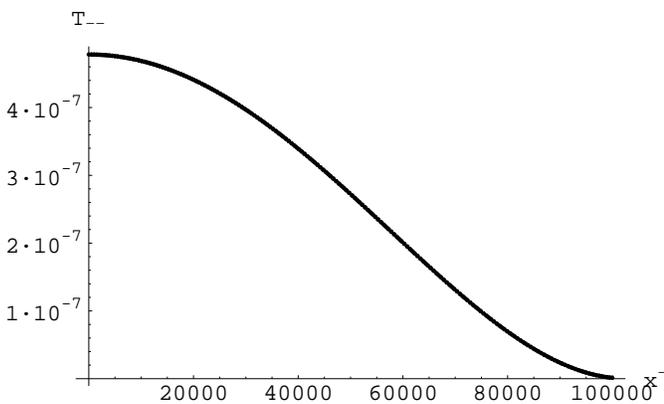}
\caption{$\tilde{T}_{--}$, the stress-energy of an affine null
  observer in physical spacetime $ds^2=-(e^{2\rho}/\sqrt{\phi})dx^+dx^-$,
  shown at points along the fixed radius
  $\pht =100$.  This positive outward flux goes to zero as the black hole
  returns to extremality. [$l=1,q=100;\Delta m=.0002;
  N\hbar=25\pi;\phi_0=2500$]}
\label{tmm}
\end{figure}
\begin{figure}[htbp]
\includegraphics[width=3.5in]{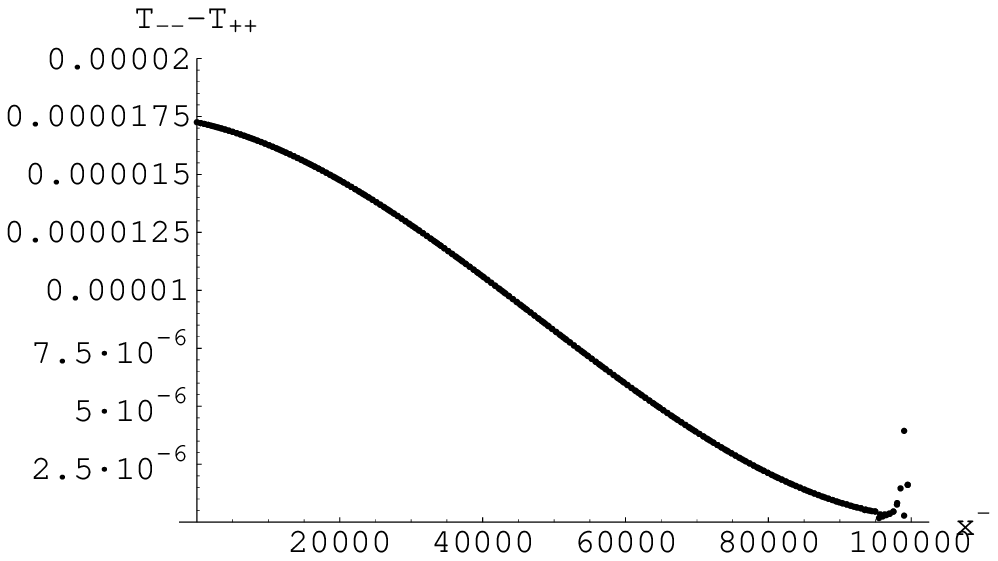}
\caption{$\tilde{T}_{--}-\tilde{T}_{++}$ , the stress-energy of an
  affine null observer in physical spacetime $ds^2=-(e^{2\rho}/\sqrt{\phi})dx^+dx^-$,
  shown at points along the fixed radius $\pht =100$.  This quantity is related to
  the differential ADM mass of the black hole.  The area under the curve corresponds
  to the finite shock mass of the black hole.  [$l=1,q=100;\Delta m=.0002;
    N\hbar=25\pi;\phi_0=2500$]}
\label{tmmmtpp}
\end{figure}
We consider the behavior in the  weak back-reaction
regime, where $N\hbar/(24\pi q^2)\ll 1$, where the adiabatic
approximation should be valid.  Far outside of the black
hole, closer to the AdS boundary, when $\phi-\phi_0 \gg
\phi_h-\phi$ (using $\phi_h$ to denote the radius at the horizon,
$\pt_+\phi=0$), while still within the validity of the
near-horizon approximation ($\phi-\phi_0\ll 1$) the flux in and out
will have a more physically intuitive interpretation.  We hope
that since the contours of $\phi$ we consider are very close to
the AdS boundary, they represent sufficiently well the behavior
that occurs at ``infinity,'' without actually leaving the
near-horizon region of our calculations.  For example, consider
the contour depicted in Fig.~\ref{phicontour}.  Here, $\phi >
\phi_0$,  yet it is small enough to be consistent with the
previous approximation, $\pht/\phi_0 \ll 1$.  We are interested in
the behavior of the physical stress-energies, as seen by an affine
observer along the inner and outer horizon, in order to test the
stability of the evaporating black hole. Fig.'s \ref{tppouter}
through Fig.~\ref{tmmouter} illustrate the values of the
stress-energy along these contours. We note from these that the
stress-energy varies smoothly throughout the evaporation process.
The stress-energy tensors for affinely parameterized observer
evaluated at each point along the fixed radial contour are shown
in Fig.~\ref{tpp} and Fig.~\ref{tmm}. In each case, the flux
approaches zero as the black hole evaporates, which is consistent
with the evaporation of a near-extremal black hole, which should
cease as the black hole returns to its extremal state.  In order
to make physical sense of the quantities, it may be useful to look
at the difference $\tilde{T}_{--} -\tilde{T}_{++}$,
Fig.~\ref{tmmmtpp}.  This is where we can have a reasonable
interpretation of what is going on, as the difference will
represent the net flux going through a surface of constant
$\phi$. The net flux going through a surface of fixed radius is
positive and vanishes away with time. The integral of this
quantity along the contour would give us the ADM mass, defined at
spatial infinity outside of a black hole. The indication then is
that the shock mass evaporates away, returning the black hole to
its extremal state. . We observe that the differential black hole
ADM mass increases for a bit, before decreasing down close to
zero. The temporary rise in mass, before dying down may be
consistent with observations by \cite{navlong}. Nevertheless, the
important quantity is its integral.

\begin{figure}[htbp]
\includegraphics[width=3.5in]{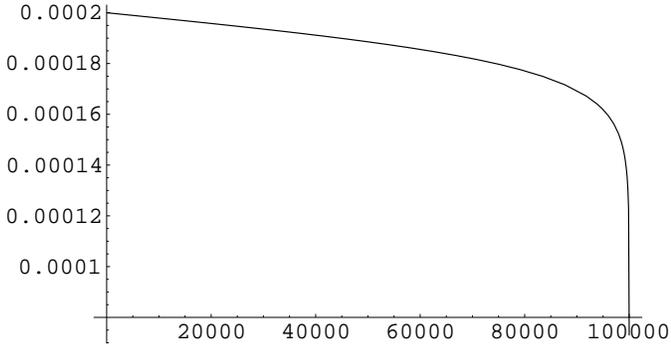}
\caption{The Bondi mass, evaluated for the conformally rescaled
  spacetime $ds^2=-e^{2\rho}dx^+dx^-$, does not indicate significant
  evaporation until close to the very end of black hole evaporation.
  This plot overshoots the zero mass point when the semiclassical
  approximation breaks down at the endpoint of evaporation.
  [$l=1,q=100; \Delta m=.0002;N\hbar=75\pi;\phi_0=2500$]}
\label{bondimass}
\end{figure}

We can also consider the Bondi mass.  This is generally defined at
future null infinity $x^+ \rightarrow \infty$, giving $m(x^-)$.
The AdS boundary of these solutions makes it difficult to use this
definition.  However, as discussed in \cite{navlong,Mann,navmass},
because $T_{--}$ is chosen everywhere to be zero, it is possible
to define a Bondi mass $m(x^+)$ for all $x^-$ with
\be
\label{bondim}
m_S(x^+)=m_{S0} - 2l\int dx^+ e^{-2\rho} \pt_- \pht T_{++} \;.
\ee
We want to verify that $\pt_-m_S=0$.  By applying the partial
derivative with respect to $x^-$ to the above we obtain
\be
\pt_- m_S(x^+)=-2l\int dx^+ e^{-2\rho} T_{++}
\left[-2\pt_-\rho\pt_-\pht+\pt^2_- \pht\right] \;,
\ee
where we used the fact that $\pt_-T_{++}=0$.  Further
manipulation, using the equations of motion, gives
\be
\pt_- m_S(x^+)=l\int dx^+
e^{-2\rho}T_{++}(T^f_{--}-\frac{N\hbar}{12\pi}t_- (x^-)) \;.
\ee
The bracketed term, then, must be constrained to zero for the mass
formula to be valid, which is indeed the case for these solutions.
A calculation of this mass using the above derived values yields
to first order in $N\hbar$
\be
m_S(x^+)=\Delta m - \frac{N\hbar}{12\pi}\sqrt{\frac{\Delta
m}{2lq^3}}{\rm arctanh}\left((x^+ -x^+_0)\sqrt{\frac{\Delta
m}{2lq^3}} \right)
  \;.
\ee
A plot of $m_S(x^+)$ (Fig.~\ref{bondimass}) shows that evaporation
occurs slowly until close to the meeting of the horizons, at which
point the mass significantly drops.  The mass evaporates to zero
at the same value of $x^+$ where the outer apparent horizon
recedes back to the extremal radius.  That is to say, when
$m_S(x^+_f)=0$, $r_0(x^+_f)=r_+(x^+_f)$.  The overshooting of the
zero-point suggests that the evaporation does not end once
extremality is reached.  However, the semiclassical description of
black hole radiation is applicable only as long as
\be
\left| T\left(\frac{\pt T}{\pt m_S}\right)\right| \ll |T|~,
\ee
where $T$ is the black hole temperature.  The temperature
fluctuations must remain small compared to the temperature itself
\cite{wilczetc}.  This means that our solutions no longer describe
the evolution of the black hole once it has returned to its
extremal state.  We can only trust our results up to $m_S(x^+)=0$,
when the outer horizon, $r_+$ and the inner horizon, $r_-$ meet
again at $r_0$.
\section{Dynamical Boundary Solutions}
\begin{figure}[htbp]
\includegraphics[width=3.5in]{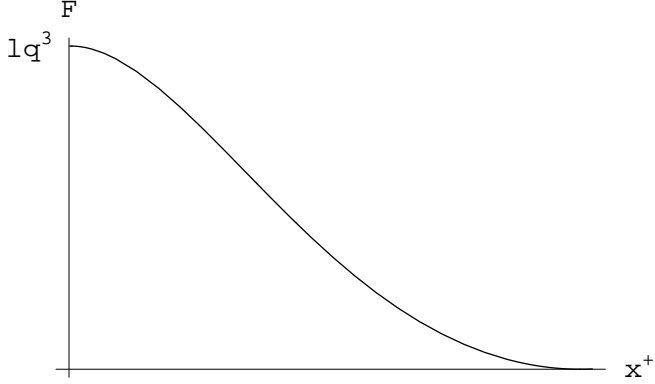}
\caption{A plot of the behavior of the function, $F(x^+)$, which
approaches zero with increasing $x^+$. $F=0$ signals the endpoint
of evaporation, which occurs at the AdS boundary, $x^-=x^+$,
representing timelike infinity.}
\label{Fofx}
\end{figure}

The other case considered by Fabbri, Navarro, and Navarro-Salas
\cite{nav2,navlong} involves a questionable choice of boundary
conditions. Here, instead of sending in a shock mass and observing
the evolution of the black hole, the shock mass is permitted
``quantum corrections'' and allowed to behave dynamically.  That
is, $m_S(v)$, the shock mass, is now allowed to vary with $v$,
instead of being expressed simply by $\Delta m \Theta(v-v_0)$
function.  In this case,  the solutions for $\pht$ cannot be found
analytically.  As stated before, it is required that $v=v(x^+)$,
or more conveniently, we choose $v(x^+)$ to be of the form
(\ref{dvdxp}).  Repeating the results following (\ref{dvdxp}), and
once again gauge fixing $\rho$ so that equation (\ref{rho}) still
holds, we see that
\be
-2\pt^2_+\pht + 4 \pt_+ \rho \pt_+ \pht =
-F'''(x^+)=T^f_{++}-\frac{N\hbar}{12\pi}t_+(x^+)\;.
\ee
From the Schwarzian derivative we have
\be
t_+(x^+)=\frac{1}{2}\{v,x^+\}=\frac{1}{2}\left(\frac{1}{2}\frac{F'(x^+)^2}{F(x^+)^2}-
\frac{F''(x^+)}{F(x^+)}\right) \;.
\ee
Thus
\be
\label{initstr}
 T^f_{++}=\Delta m\delta(x^+-x^+_0)
=-F'''(x^+)+\frac{N\hbar}{24\pi}\left(\frac{1}{2}\left(\frac{F'(x^+)}
    {F(x^+)}\right)^2-\frac{F''(x^+)}{F(x^+)}\right) \;.
\ee
Continuity of $\pht$ requires that
\bea\label{bc1}
F(x^+_0)&=&lq^3 \;,\\
F'(x^+_0)&=&0 \; .
\eea
It follows further from (\ref{initstr}) that
\be\label{bc3}
F''(x^+_0)=-\Delta m \;.
\ee
It is now possible to numerically solve for $F(x^+)$, using
(\ref{initstr}) and the boundary conditions (\ref{bc1}-\ref{bc3}).
Typical behavior for $F(x^+)$ can be seen in Fig.~\ref{Fofx}.

We set
\be
\left(-2\frac{\pht^2}{q^3}+l\Delta m\right)(\pt_+v(x^+))^2+2l\pt_+
   \pht\pt_+v(x^+)=0 \;,
\ee
which is required in order to eliminate off-diagonal components in
(\ref{confmetric}), when making the coordinate transformations.
This allows us to solve for
\be\label{bondi2}
m_S(x^+)=\frac{F'^2(x^+)-2F(x^+)F'(x^+)}{2lq^3} \;.
\ee
Using (\ref{initstr}) this can be rewritten as
\be
m_S(x^+)=\frac{24\pi}{N\hbar lq^3}F^2(x^+)F'''(x^+) \;.
\ee
In a less straightforward manner, we could have also used
(\ref{partialmphi}) and (\ref{rho}) to evaluate the mass based on
the previous definition (\ref{bondim}).
\bea
\pt_+m_S(x^+)&=&-\frac{F(x^+)F'''(x^+)}{lq^3}\\
    &=&-\frac{F(x^+)F'''(x^+)+F(x^+)F''(x^+)-F(x^+)F''(x^+)}{lq^3}\\
    &=&\pt_+ \left(\frac{24\pi F(x^+)^2F'''(x^+)}{N\hbar lq^3}\right) \;,\\
\eea
arriving at (\ref{bondi2}).  We can see now that
\be
\pt_+m_S(x^+)=-\frac{N\hbar}{24\pi}\frac{m_S(x^+)}{F(x^+)}\; .
\ee
Multiplying by $dx^+/dv$
\be
\pt_v m_S(v)=-\frac{N\hbar}{24\pi lq^3}m_S(v)\; ,
\ee
so that
\be
m_S(v)=\Delta m
e^{-\frac{N\hbar}{24\pi}\frac{v-v_0}{lq^3}}\Theta(v-v_0)\; .
\ee
Using a complementary method,it is also possible to define instead
a late time Bondi mass \cite{navcomp1,navcomp2}, which behaves as
\be
\pt_u m_S(u)=-\frac{N\hbar}{24\pi lq^3}m_S(u)\;.
\ee
We can now use the fact that the four-dimensional stress-energy
tensor can be related back to the rate of change of the mass
\cite{navlong,vaidya},
\be
T^{(4)}_{vv}=\pt_v m_S(v)=\Delta m
\delta(v-v_0)-\frac{N\hbar}{24\pi lq^3}m_S(v)\Theta (v- v_0)~,
\ee
to see that there is a negative flux of energy being sent in to
reduce the black hole mass \cite{navcomp1}.  Generally, however,
it does not make sense to allow the shock mass being sent in to
vary as a function of $x^+$ for all values of $x^-$, including
those corresponding to very large distances away, where quantum
effects due to an ``infinitely'' far black hole ought to be
negligibly small.  Since $m_S(v)$ now behaves as shown above, as
opposed to simply $m_S(v)=\Delta m \Theta(v-v_0)$, then this
solution corresponds to sending in negative mass after the initial
$\Delta m$.  This is an artifact of the boundary condition
requirement which forces $T_{--}=0$ everywhere, so that basically
the black hole evaporation occurs through flux being sent in from
infinity outside.  The previous boundary condition, however, was
more natural, since the diminishing mass was arrived at, rather
than put in by hand. Nonetheless, it may be argued that the
difference between these two boundary conditions is minute. In
both cases, the same general phenomena is occurring; $T_{--}=0$
and $T_{++}<0$ bring about the evolution of the black hole.

The stress-energy tensor can again be calculated for the affinely
null coordinates.  We proceed in the same manner as before,
evaluating $\tilde{T}_{++}$ and $\tilde{T}_{--}$ for the physical
metric (\ref{physmet}).
\begin{figure}[htbp]
\includegraphics[width=3.5in]{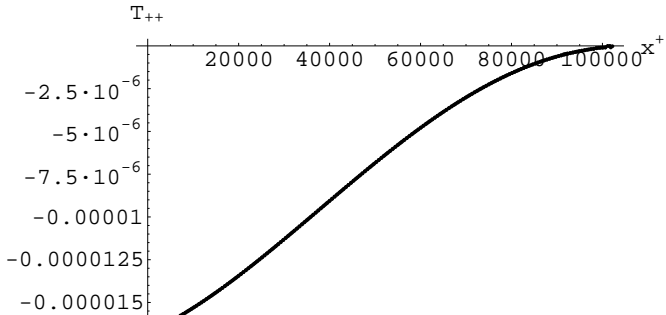}
\caption{$\tilde{T}_{++}$, the stress-energy of an affine null
  observer in physical spacetime $ds^2=-e^{2\rho}/\sqrt{\phi}dx^+dx^-$,
  evaluated at points along the fixed radius $\pht =100$.  Again, the flux is
  negative, indicating that negative mass is being sent in as a result of the
  boundary conditions.  This flux goes to zero as the black hole evaporation draws
  to an end. [$l=1,q=100; \Delta
  m=.0002;N\hbar=75\pi;\phi_0=2500$]}
\label{tpp2}
\end{figure}
\begin{figure}[htbp]
\includegraphics[width=3.5in]{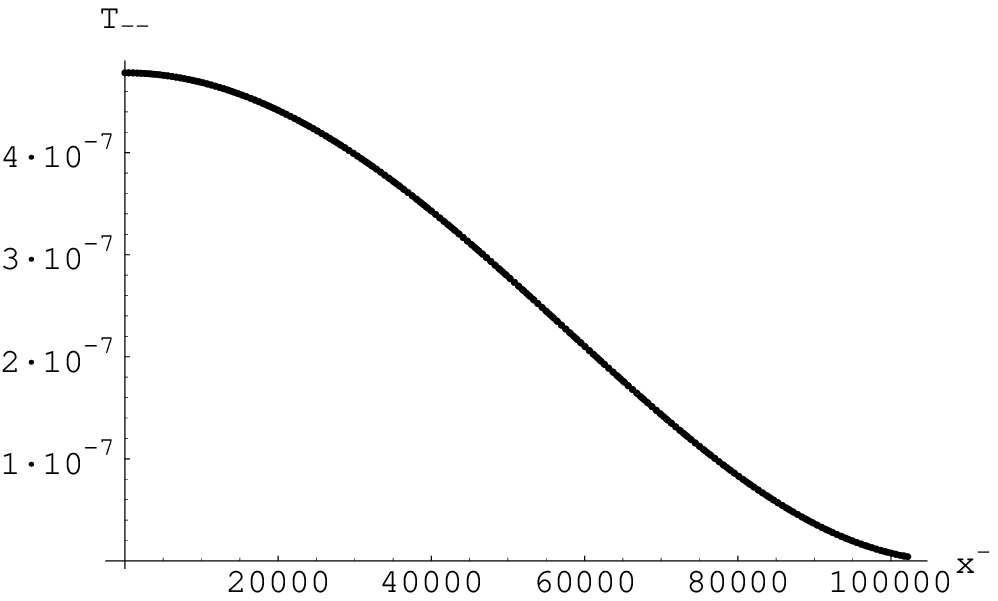}
\caption{$\tilde{T}_{--}$, the stress-energy of an affine null
  observer in physical spacetime $ds^2=-e^{2\rho}/\sqrt{\phi}dx^+dx^-$,
  evaluated at points along the fixed radius $\pht =100$.  This outward positive
  flux goes to zero as the black hole returns to extremality.  [$l=1,q=100;
  \Delta m=.0002;N\hbar=75\pi;\phi_0=2500$]}
\label{tmm2}
\end{figure}
\begin{figure}[htbp]
\includegraphics[width=3.5in]{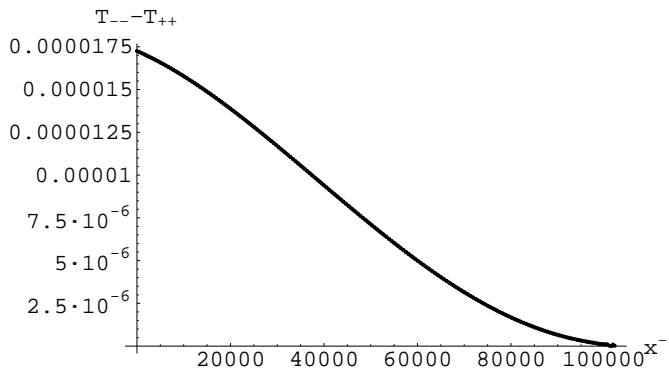}
\caption{$\tilde{T}_{--}-\tilde{T}_{++}$ , the stress-energy of an
  affine null observer in physical spacetime $ds^2=-e^{2\rho}/\sqrt{\phi}dx^+dx^-$,
  shown at points along the fixed radius $\pht =100$.  This quantity is related to
  the differential ADM mass of the black hole.  The area under the curve corresponds
  to the finite shock mass of the black hole.  The boundary conditions were chosen so
  that this quantity goes to zero monotonically.
  [$l=1,q=100; \Delta m=.0002;N\hbar=75\pi;\phi_0=2500$]}
\label{tmmmtpp2}
\end{figure}
We see from the stress-energy tensors, $\tilde{T}_{++}$, in
Fig.~\ref{tpp2} that there is indeed a negative flux of energy
entering the black hole.  In order to bring about exponential
decay of the shock mass, it was necessary to send in negative mass
to bring it down.   $\tilde{T}_{--}$, (Fig.~\ref{tmm2}) behaves
also as expected.  There is a flux of energy being emitted from
the black hole which approaches zero as the black hole evaporates
away.  The differential ADM mass, as represented by
$\tilde{T}_{--}-\tilde{T}_{++}$ (Fig.~\ref{tmmmtpp2}) decays down
to zero, as a result of the combined effect of the negative energy
being sent in and the natural black hole emission.
\section{Conclusion}

We have considered toy models for semiclassical charged black hole
evaporation with back reaction included.
The calculations of the physical stress-energy tensor
of a freely falling observer, allowed us to verify that an observer
at a fixed distance outside of the black hole
perceives a flux decaying to zero, as the
black hole evaporates away the injected shock mass and returns to the
extremal state.  The affine stress energy
also varies smoothly across the inner and outer apparent horizons,
suggesting that there is no buildup of energy in places which
would have undermined the validity of the semiclassical solutions.
Hence, while in these solutions the black hole evaporation is
largely dictated by the negative flux of energy entering the black
hole, the vacuum choices discussed do indeed yield a picture
consistent with charged black hole evaporation in string theory
\cite{lowe}
without encountering the
singularities feared by \cite{jacobson}.  Nevertheless, there are
some questions and difficulties of interpretation that would be
better answered by working in an asymptotically Minkowskian
spacetime. Most notably, we are troubled by the apparent
arbitrariness of the imposed boundary conditions. It would be more
natural to choose a vacuum in a spacetime that corresponds to what
an observer infinitely far from the black hole might observe.  In
addition, the evaporation of the black hole in these solutions is
inherently dictated \`{a} priori by $t_{\pm}(x^{\pm})$. It would
be more satisfying if quantum effects did not have to be
predicated, but rather were naturally manifested as the solutions
evolved forward in time. More physically motivated boundary
conditions render the equations of motion less easily solvable,
making it necessary to solve a set of coupled partial differential
equations instead of the ordinary differential equations used
here.  This would have to be done numerically.  This work is
currently being completed.  Preliminary results, however, seem to
be consistent with those presented here \cite{diba}.

\bigskip
\goodbreak
\centerline{\bf Acknowledgements}
\noindent
This research is supported in part by DOE
grant DE-FE0291ER40688-Task A.

\end{document}